\documentclass[aps,pre,reprint,amsmath,amssymbols,superscriptaddress,
nofootinbib]{revtex4-1}

\usepackage{graphicx}
\usepackage{bm}
\usepackage{color}
\usepackage{soul}

\definecolor{darkBlue}{rgb}{0,0,1}
\definecolor{darkRed}{rgb}{0.5,0,0}
\definecolor{darkGreen}{rgb}{0,0.5,0}

\begin{document}


\title{Uniaxial and biaxial structures in the elastic Maier-Saupe model}

\author{A. Petri}
\affiliation{CNR - Istituto dei Sistemi Complessi, Dipartimento di Fisica, Universit\`{a} Sapienza, Roma, Italy}
\email{alberto.petri@isc.cnr.it}
\author{D. B. Liarte}
\affiliation{Laboratory of Atomic and Solid State Physics, Cornell University, Ithaca, NY 14853-250, USA}
\email{dl778@cornell.edu}
\author{S. R. Salinas}
\affiliation{Instituto de F\'{\i}sica, Universidade de S\~{a}o Paulo, S\~{a}o Paulo, SP, Brazil}
\email{ssalinas@if.usp.br}




\date{\today}

\begin{abstract}
We perform statistical mechanics calculations to analyze the global phase
diagram of a fully-connected version of a Maier-Saupe-Zwanzig lattice model
with the inclusion of couplings to an elastic strain field. We point out the
presence of uniaxial and biaxial nematic structures, depending on temperature
$T$ and on the applied stress $\sigma$. Under uniaxial extensive tension,
applied stress favors uniaxial orientation, and we obtain a first-order
boundary, along which there is a coexistence of two uniaxial paranematic
phases, and which ends at a simple critical point. Under uniaxial compressive
tension, stress favors biaxial orientation; for small values of the coupling
parameters, the first-order boundary ends at a tricritical point, beyond which
there is a continuous transition between a paranematic and a biaxially ordered
structure. For some representative choices of the model parameters, we obtain
a number of analytic results, including the location of critical and
tricritical points and the line of stability of the biaxial phase.
\end{abstract}


\maketitle


\section{Introduction}

Nematic elastomers are an interesting class of soft-matter systems, with an
interplay of the effects of nematic order and of elastic properties of
rubber materials \cite{Warner2003}. At fixed and sufficiently low
temperatures, elastomer systems display stress-strain curves with a plateau
that resembles a liquid-gas transition. Since the pioneering work of de Gennes and
collaborators \cite{deGennes1975,deGennesProst1993,deGennesOkumura2003},
the couplings between elastic and orientational degrees of freedom have been
recognized as the essential ingredient to account for the behavior of elastomers.
More recently, it has been argued that it is also important to introduce quenched
random fields, which are supposed to smooth the first-order transition, and which
are related to the phase history of the cross-links of the polymer network
\cite{Fridrikh1997,Yu1998,Selinger2002,Selinger2004,Fridrich2006,Petridis2006,
Zhu2011}. In a recent work, some of us performed mean-field
calculations for a basic Maier-Saupe lattice model, which was supplemented by
the addition of an elastic strain field and of random fields \cite{Liarte2011,
Liarte2013}. At the qualitative level, it has been possible to account
for many of the experimental properties of uniaxial nematic elastomers. We
were then motivated to revisit this elementary lattice model, with the
addition of a simple form of elastic coupling, and with no random fields. The
fully-connected version of this model system, as we define in Section 2, is
amenable to detailed and standard statistical mechanics calculations
\cite{doCarmo2011,Liarte2014}. We show that the introduction of a
coupling to an external strain field is sufficient to provide a mechanism to
change and depress the standard first-order nematic transition, and make contact
with the plateaux in the stress-strain curves of nematic elastomers. In Section 3,
we obtain a number of analytic results for the stress-temperature phase
diagram of some simple representative cases. We show the existence of a rich
behavior, including critical and tricritical points and the continuous
transition to a biaxial nematic structure.

As in the original phenomenological treatment of de Gennes and collaborators
\cite{deGennesOkumura2003}, the external stress in the elastic nematic model
system plays a similar role as an applied magnetic (electric) field in the
corresponding uniaxial nematic system. A magnetic field couples to the nematic
order via a constant $\chi_{a}$ that can be either positive or
negative~\cite{deGennesProst1993}. Positive values of this susceptibility
($\chi_{a}>0$) correspond to positive tensions (extension) in the elastic
model. In the phase diagrams, either in terms of stress or in terms of fields
and temperature, there is a line of first-order transitions that ends at a
simple critical point~\cite{Petri2016}. Negative values of the susceptibility
($\chi_{a}<0$) lead to a field-induced biaxial phase.

It is interesting to remark that Ye, Lubensky and coworkers
have considered a minimal model \cite{Ye2007,Ye2009}, at the
phenomenological level, which provides a robust
description of the elastic semisoft response of nematic elastomers. In
addition to the uniaxial strain field, these authors also add a transverse
strain component, which might as well be produced in the cross-linking
K\"{u}pfer-Finkelmann procedure of preparing elastomer samples
\cite{Kuepfer1991}. They point out that this minimal model is analogous to
the formulation of a nematic system in the presence of crossed electric and
magnetic field, which also leads to a rich phase diagram, with critical and
tricritical behavior, and the possibility of existence of a biaxial phase. In the
present work, however, we consider a minimal lattice model, and keep the
restriction to an elastic strain field along the uniaxial direction.

\begin{figure}
[ptb]
\begin{center}
(a)
\vspace{0.2cm}

\includegraphics[width=0.85\linewidth]{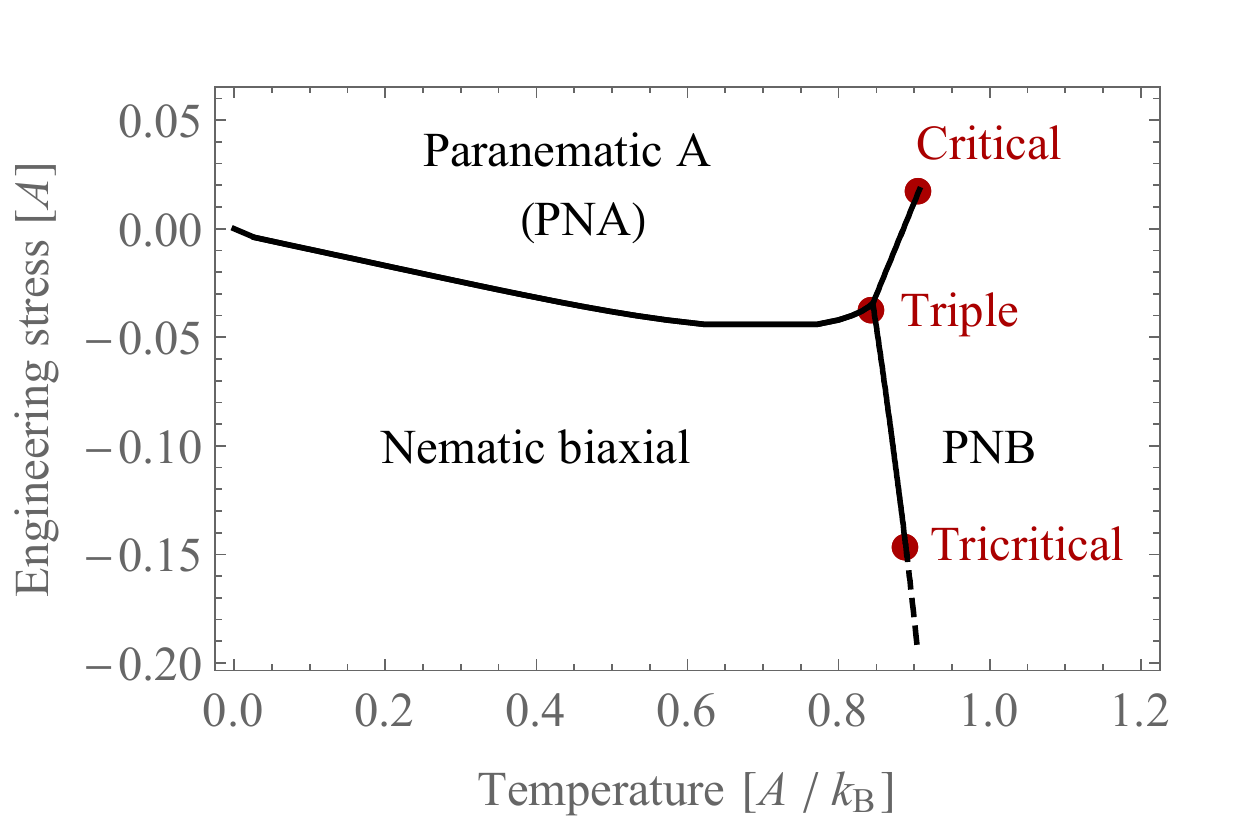}%
\vspace{0.2cm}

(b)
\vspace{0.2cm}

\includegraphics[width=0.85\linewidth]{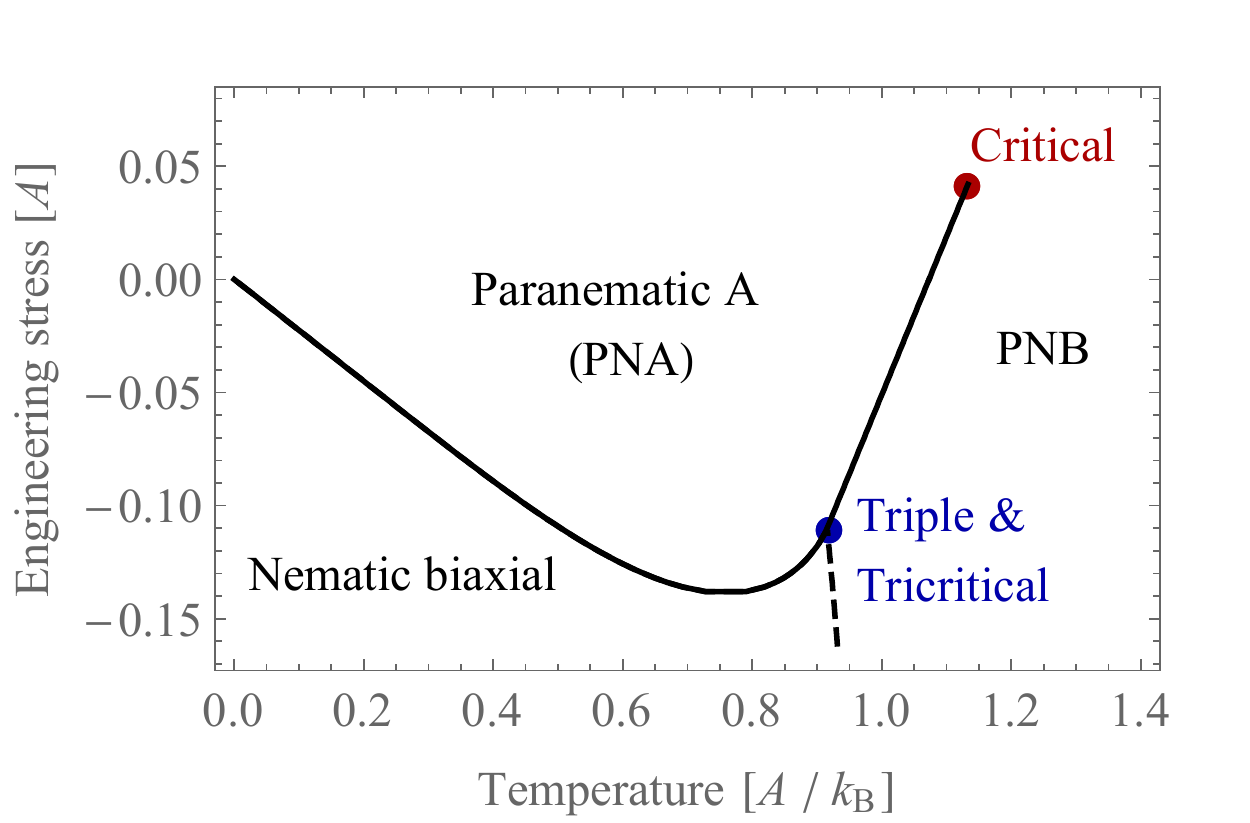}%
\caption{Sketch of the stress-temperature phase diagram of a uniaxial nematic
system for small (a) and large (b) values of the
coupling parameters. Solid and dashed lines correspond to first and
second-order transitions. We indicate critical and tricritical points.}%
\label{fig:PhaseDiagram}
\end{center}
\end{figure}

In the elastic model of the present work, negative tensions (compression) may
lead to a stress-induced biaxial phase. In figure \ref{fig:PhaseDiagram}a, for
sufficiently weak and typical values of the coupling parameters, we draw a sketch
of a representative phase diagram of a uniaxial nematic system in terms of
applied stress (along the uniaxial direction) and temperature. For larger values of
the coupling parameters, the tricritical point collapses at the first-order
boundary, and the topology of this diagram may be drastically changed (figure
\ref{fig:PhaseDiagram}b).

\section{The elastic Maier-Saupe model}

In the Maier-Saupe approach to the nematic transitions, we consider a lattice
of $N$ sites and write the energy%
\begin{equation}
\mathcal{H}_{MS}=-A\sum_{\left(  i,j\right)  }\sum_{\mu,\nu=x,y,z}S_{i}%
^{\mu\nu}S_{j}^{\mu\nu},
\end{equation}
where $A>0$ is an interaction parameter, the first sum is over
nearest-neighbor pairs of lattice sites, and the local \textquotedblleft
quadrupolar\textquotedblright\ degrees of freedom are given by
\begin{equation}
S_{i}^{\mu\nu}=\frac{1}{2}\left(  3n_{i}^{\mu}n_{i}^{\nu}-\delta_{\mu\nu
}\right)  ,
\end{equation}
where $\overrightarrow{n}_{i}$ is the unit director associated with a
nematogenic molecule at site $i=1,2,...,N$, and $\delta_{\mu\nu}$ is the
Kronecker delta.

In the mean-field calculations of this work, we assume a fully-connected
version of this Maier-Saupe model, given by the energy%
\begin{equation}
E_{MS}=-\frac{A}{N}\sum_{1\leq i<j\leq N}\sum_{\mu,\nu=x,y,z}S_{i}^{\mu\nu
}S_{j}^{\mu\nu},\label{ems}%
\end{equation}
where the first sum is over all pairs of lattice sites, and the parameter $A$
is suitably scaled to preserve the existence of the thermodynamic limit.
We further assume that the local directors $\overrightarrow{n}_{i}$ are
restricted to the Cartesian axes,%
\begin{equation}
\overrightarrow{n}_{i}=\left\{
\begin{array}
[c]{c}%
\left(  \pm1,0,0\right)  ,\\
\left(  0,\pm1,0\right)  ,\\
\left(  0,0,\pm1\right)  ,
\end{array}
\right. \label{directors}%
\end{equation}
which leads to a three-state statistical model. This choice of discrete
orientational degrees of freedom, which resembles an approximation used by
Zwanzig to treat the Onsager model for the nematic transition, leads to a
problem that is amenable to quite simple statistical-mechanics calculations
(and to simple simulations as well). Moreover, it is known that this
elementary MSZ model leads to essentially the same qualitative results as
obtained from slightly more involved calculations for the original Maier-Saupe
model with a continuous distribution of orientational degrees of freedom
\cite{Liarte2014,Nascimento2016}. We should keep in mind, however, that
the continuous symmetry may be strictly necessary to describe some special
features of these systems, as the soft transitions observed in liquid-crystal
elastomers \cite{Liarte2013}. It is possible that the soft/semisoft response
of nematic elastomers depletes the biaxial nematic state.

According to the notation of our previous article \cite{Liarte2011}, the
energy of a microscopic configuration of the elastic Maier-Saupe lattice model
is written as a sum of three terms,%
\begin{equation}
E=E_{MS}+E_{elastic}+E_{coupling},\label{energy}%
\end{equation}
where $E_{MS}$ is given by eq. (\ref{ems}). Here we ignore the
quenched random fields originating in the aligning stresses that are generated
at the time of cross-linking. Hence, positive stresses lead to a nematic director
that is aligned with the axis of deformation.

The theory of rubber elasticity, as developed by Warner and
Terentjev \cite{Warner2003}, leads to the elastic free energy per site of a
lattice polymer,%
\begin{equation}
f_{rub}=\frac{1}{2}\mu_{s}\left(  \lambda^{2}+\frac{2}{\lambda}\right)  ,
\end{equation}
where $\mu_{s}>0$ is a linear shear modulus and $\lambda$ is the distortion
factor of a uniaxial, volume-preserving, deformation. In order to keep the
calculations as simple as possible, and restrict the analysis to an elementary
model that is still capable of qualitatively accounting for the effects of
strain, we expand $f_{rub}$ about the minimum, $\lambda=1$, and keep quadratic
terms only. We then discard an additive term, $\left(  3/2\right)  \mu_{s}%
N$, and write the elastic energy,%
\begin{equation}
E_{elastic}=\frac{3}{2}\mu_{s}N\left(  \lambda-1\right)  ^{2}.\label{eelastic}%
\end{equation}

We now discuss the couplings between orientational degrees of freedom and a
uniform external strain along a certain direction. According to previous work
\cite{Liarte2011}, we introduce a global tensor,%
\begin{equation}
M_{\mu\nu}=\frac{1}{2}\left(  3m^{\mu}m^{\nu}-\delta_{\mu\nu}\right)
,\label{Mmunu}%
\end{equation}
which characterizes a global uniaxial deformation, where the stress is applied
along the unit vector $\overrightarrow{m}$ (which is distinct from the local
nematic directors). If we use the Warner-Terentjev theory \cite{Warner2003},
and keep the dominant contribution near the minimum $\lambda=1 $, the coupling
term of the energy is written as%
\begin{equation}
E_{coupling}=-B\sum_{i=1}^{N}\sum_{\mu,\nu}M_{\mu\nu}S_{i}^{\mu\nu
},\label{ecoupling}%
\end{equation}
with%
\begin{equation}
B=\mu_{s}\delta\left(  \lambda-1\right)  ,\label{B}%
\end{equation}
where the parameter $\delta$ gauges the strain anisotropy (and is usually
assumed to be positive).

We emphasize that rubber elasticity is a primarily entropic phenomenon.
Polymer chains optimally arrange themselves according to high-entropy
configurations instead of configurations that lower an interatomic or
molecular potential. Therefore, the elastic and coupling terms in equation
(\ref{energy}) come from a degeneracy factor associated with the
configurations of the polymer network, and should be regarded as effective
energy terms \cite{Liarte2011,Liarte2013}. Also, the linear chain
modulus is usually taken as proportional to temperature, so in this work we
assume that%
\begin{equation}
\mu_{s}=n_{s}k_{B}T,\label{mus}%
\end{equation}
where $T$ is the temperature, $k_{B}$ is the Boltzmann constant, and $n_{s}$
is the relative number of polymer strands.

The thermodynamic properties of this elastic MSZ model can be obtained from a
canonical partition function in the stress ensemble, which is given by%
\[
Y=Y\left(  T,\sigma,N,\left\{  M_{\mu\nu}\right\}  \right)  =\int d\lambda
\exp\left[  N\beta\sigma\lambda\right]  \times
\]%
\begin{equation}
\times\sum_{\left\{  \overrightarrow{n}_{i}\right\}  }\exp\left[
-\beta\left(  E_{MS}+E_{elastic}+E_{coupling}\right)  \right]  ,
\end{equation}
where $\beta=1/k_{B}T$, $\sigma$ is the global engineering stress, and the sum
is over all configurations of the local nematic directors. Taking into account
equations (\ref{ems}), (\ref{eelastic}), and (\ref{ecoupling}), and discarding
irrelevant terms in the thermodynamic limit, we write%
\begin{align}
Y &= \int d\lambda\exp\left[  N\beta\sigma-\frac{3}{2}N\beta\mu_{s}-\frac{3}%
{2}N\beta\mu_{s}\,\left(  \lambda-1\right)  ^{2}\right]
\nonumber \\
& \quad \times \sum_{\left\{
\overrightarrow{n}_{i}\right\}  }\exp\left[  -\beta E_{eff}\right]  ,
\end{align}
with%
\begin{equation}
E_{eff}=-\frac{A}{2N}\sum_{\mu,\nu}\left(  \sum_{i=1}^{N}S_{i}^{\mu\nu
}\right)  ^{2}-B\sum_{i=1}^{N}\sum_{\mu,\nu}M_{\mu\nu}S_{i}^{\mu\nu},
\end{equation}
where $A>0$ is a constant dimensional parameter, and the expression of $B$ is
given by eqs. (\ref{B}) and (\ref{mus}),%
\begin{equation}
B=n_{s}k_{B}T\delta\left(  \lambda-1\right) \label{BB}.%
\end{equation}

We now resort to well-known techniques of statistical mechanics. If we use a
set of Gaussian identities to deal with the quadratic terms, and change to
more convenient variables, it is straightforward to write%
\begin{align}
& \sum_{\left\{  \overrightarrow{n}_{i}\right\}  }\exp\left[  -\beta
E_{eff}\right] =\int\left[  dQ_{\mu\nu}\right] 
\nonumber \\ & \quad \times
\exp\left[  -\sum_{\mu,\nu
}\frac{\beta AN}{2}Q_{\mu\nu}^{2}\right]  \,\left(  Z_{1}\right)  ^{N},
\end{align}
with the short-hand notation%
\begin{equation}
\int\left[  dQ_{\mu\nu}\right]  \left(  ...\right)  =%
{\displaystyle\prod\limits_{\mu,\nu}}
\int\left(  \frac{\beta AN}{2\pi}\right)  ^{1/2}dQ_{\mu\nu}\left(  ...\right)
,
\end{equation}
where $Z_{1}$ is a single-particle partition function,%
\begin{equation}
Z_{1}=\sum_{\overrightarrow{n}}\exp\left\{  \beta\sum_{\mu,\nu}\left[
AQ_{\mu\nu}+BM_{\mu\nu}\right]  S_{\mu\nu}\right\}  .
\end{equation}
Performing the sum over the nematic directors according to the six
possibilities of eq. (\ref{directors}), we finally have%
\begin{align}
Z_{1} &=2\exp\left[  -\frac{\beta A}{2}\sum_{\mu}Q_{\mu\mu}\right]  \sum_{\mu
}\exp\left[  \frac{3}{2}\beta AQ_{\mu\mu}
\right. \nonumber \\ & \quad \left.
+\frac{3}{2}\beta BM_{\mu\mu}\right].
\end{align}

We now use these expressions to write%
\begin{equation}
Y=\int d\lambda\int\left[  dQ_{\mu\nu}\right]  \exp\left[  -\beta
Ng_{f}\right]  ,
\end{equation}
where the free-energy functional is given by%
\begin{align}
g_{f}&=-\sigma\lambda+\frac{3}{2}\mu_{s}+\frac{3}{2}\mu_{s}\left(
\lambda-1\right)  ^{2}+\frac{1}{2}A\sum_{\mu,\nu}Q_{\mu\nu}^{2}
\nonumber \\ & \quad
+\frac{1}{2}A\sum_{\mu}Q_{\mu\mu}
-\frac{1}{\beta}\ln2-\frac{1}{\beta}\ln\left\{  \sum_{\mu}
\right. \nonumber \\ & \quad \left.
\exp\left[ \frac{3}{2}\beta AQ_{\mu\mu}
+\frac{3}{2}\beta BM_{\mu\mu}\right]  \right\}
.\label{gf}%
\end{align}
It should be remarked that $g_{f}$ is a function of temperature $T$, global
external stress $\sigma$, and the components of the tensor $M_{\mu\nu}$, as
well as of the strain distortion $\lambda$ and the components of the tensor
order parameter $Q_{\mu\nu}$. The thermodynamic free energy per site,
$g=g\left(  T,\sigma \right)  $, comes from a
saddle-point calculation, which amounts to a minimization of $g_{f}$ with
respect to $\lambda$ and $\left\{  Q_{\mu\nu}\right\}  $,%
\begin{equation}
g=g\left(  T,\sigma \right)  =\min
\limits_{\lambda,\left\{  Q_{\mu\nu}\right\}  }g_{f}\left(  T,\sigma ;\lambda,\left\{  Q_{\mu\nu}\right\}  \right)  .
\end{equation}

From the saddle-point equations, we have%
\begin{equation}
Q_{\mu\mu}=\frac{1}{2}\left(  \frac{3e_{\mu}}{\sum_{\nu}e_{\nu}}-1\right)
,\label{qmumu}%
\end{equation}
where%
\begin{equation}
e_{\mu}=\exp\left[  \frac{3}{2}\beta AQ_{\mu\mu}+\frac{9}{4}\beta Bm_{\mu}%
^{2}\right]  ,\label{emu}%
\end{equation}
and $m_{\mu}$ is a component of the unit vector $\overrightarrow{m}$
associated with the global tensor $M_{\mu\nu}$. It is easy to see that
$Q_{\mu\nu}=0$ for $\mu\neq\nu$, and that $Q_{\mu\nu}$ is a traceless tensor,%
\begin{equation}
\sum_{\mu}Q_{\mu\mu}=0.
\end{equation}
The derivative with respect to $\lambda$ leads to the remaining saddle-point
equation,%
\begin{equation}
-\frac{\sigma}{\mu_{s}}+3\left(  \lambda-1\right)  -\frac{3}{2}\delta\sum
_{\mu}m_{\mu}^{2}\,Q_{\mu\mu}=0.\label{sigma}%
\end{equation}

At this point the problem is formulated in quite general terms. We use
equations (\ref{qmumu}) and (\ref{sigma}) for obtaining the nematic order
parameter $Q_{\mu\mu}$ and the distortion $\lambda$ in terms of temperature
$T$, external stress $\sigma$, global strain $M_{\mu\nu}$, and the parameters
of the model. If there are multiple solutions, we have to search for the
absolute minima of the functional $g_{f}\left(  T,\sigma,\left\{  M_{\mu\nu
}\right\}  ;\,\lambda,\left\{  Q_{\mu\nu}\right\}  \right)  $. Several choices
of parameters and particular cases can be analyzed rather easily.

\section{Calculations for some special cases}

We now use the standard representation of the nematic tensor order parameter,%
\begin{equation}
\mathbf{Q}=\left(
\begin{array}
[c]{ccc}%
-\frac{1}{2}\left(  S+\eta\right)  & 0 & 0\\
0 & -\frac{1}{2}\left(  S-\eta\right)  & 0\\
0 & 0 & +S
\end{array}
\right)  .\label{Q}%
\end{equation}
Also, we assume that the global external strain is applied along the $z$
direction,%
\begin{equation}
\overrightarrow{m}=\left(  0,0,1\right)  .
\end{equation}

From the saddle-point equations, (\ref{qmumu}) and (\ref{sigma}), it is
straightforward to write the mean-field equations of state%
\begin{equation}
S=
1-\frac{3}%
{2+\exp{\left( 9 \, \beta \, (B + S) / 4 \right )} \, \text{sech}(3 \, \beta \, \eta / 4)},
\label{S}%
\end{equation}%
\begin{equation}
\eta=
\frac{3\, \tanh \left(3 \, \beta \, \eta / 4 \right)}%
{2+\exp{\left( 9 \, \beta \, (B + S) / 4 \right )} \, \text{sech} \left(3 \, \beta \, \eta / 4 \right)},
\label{eta}%
\end{equation}
and%
\begin{equation}
\frac{\sigma}{\mu_{s}}=3\left(  \lambda-1\right)  -\frac{3}{2}\delta
S,\label{lambda}%
\end{equation}
where $B$ is given by equation (\ref{BB}), and we are setting $A=1$ to
simplify the notation. The free energy functional can be written as%
\begin{align}
g_{f} & =-\sigma\lambda+\frac{3}{2}\mu_{s}+\frac{3}{2}\mu_{s}\left(
\lambda-1\right)^{2}+\frac{1}{2}\left(  \frac{3}{2}S^{2}+\frac{1}{2}\eta
^{2}\right)
\nonumber \\ & \quad
-\frac{1}{\beta}\ln2+\frac{3}{4}\mu_{s}\delta\left(
\lambda-1\right)  -
-\frac{1}{\beta}\ln\left[  2\exp\left(  -\frac{3}{4}\beta S\right)
\right. \nonumber \\ & \quad \left. \times
\cosh\left(  \frac{3}{4}\beta\eta\right)
+\exp\left(  \frac{3}{2}\beta
S+\frac{9}{4}\beta B\right)  \right]  .\label{gfgen}%
\end{align}

From equation (\ref{lambda}), we have%
\begin{equation}
\left(  \lambda-1\right)  =\frac{\sigma}{3\mu_{s}}+\frac{1}{2}\delta
S.\label{lambda2}%
\end{equation}
Therefore, the coupling parameter $B$ to the strain field is given by%
\begin{equation}
B=\mu_{s}\delta\left(  \lambda-1\right)  =\frac{1}{3}\sigma\delta+\frac{1}%
{2}\mu_{s}\delta^{2}S,\label{Bcomplete}%
\end{equation}
so that the strength of this coupling depends on the stress. For $\delta>0$
and $\sigma$ sufficiently large, the system prefers to align uniaxially (there
is a stable uniaxial solution, $S\neq0$ and $\eta=0$). For negative and
sufficiently large values of the stress $\sigma$, with positive strain
anisotropy, $\delta>0$, we anticipate the possibility of a biaxial arrangement
($S\neq0$ and $\eta\neq0$).

\subsection{Uniaxial transitions}

In zero stress, $\sigma=0$, we have $B=\mu_{s}\delta^{2}S/2$, so that $S=0 $
(with $\eta=0$) is a solution of the saddle point equations. At low
temperatures, however, this fully disordered solution is unstable, and there
appears a thermodynamic stable, uniaxially ordered solution, $S\neq0$ and
$\eta=0$. For $\sigma\neq0$, however, the solution $S=0$ is no longer acceptable.

In the stress-temperature phase diagram, assuming $\delta>0$, we anticipate
the existence of a line of coexistence of two distinct and uniaxially ordered
phases. Along this first-order line, as the temperature increases, there is a
decrease of the difference $\Delta S$ between the scalar order parameters
associated with the coexisting phases, which finally vanishes at a simple
critical point. This depression of $\Delta S$ is one of the hallmarks of the
behavior of elastomers \cite{Warner2003}. Given the parameters of the model,
we can draw a number of graphs of the solution $S$ versus temperature $T$ for
various values of the external stress $\sigma$ (see figures in the article by
Liarte, Yokoi, and Salinas \cite{Liarte2011}). Also, we can draw graphs of
stress $\sigma$ versus strain $e=\lambda-1$, at constant temperature, which
are shown to display characteristic plateaux, as it is experimentally observed
in nematic elastomers.

The formulation of this problem is so simple that we can resort to an earlier
analysis for the field-behavior of the Landau expansion \cite{Hornreich1985}
in order to analytically locate the critical point in the $\sigma-T$ phase
diagram. According to the usual assumptions of the mean-field approach, the
scalar order parameter $S$ in the neighborhood of the critical point may be
written as%
\begin{equation}
S=S_{c}+s,\label{s}%
\end{equation}
where $S_{c}$ is the value of $S$ at the critical point and $s$ is a small
quantity. We then insert this form of $S$ in the equation of state (with
$\eta=0$, since there is no biaxial phase), and obtain the expansion%
\begin{equation}
a+b\,s+c\,s^{2}+c\,s^{3}+...=0,\label{sexp}%
\end{equation}
where the coefficients, $a$, $b$, ..., are functions of $\sigma$, $T$, and the
parameters of the model. At the critical point, that is, with $\delta
\sigma=\delta\sigma_{c}$ and $T=T_{c}$, we should have%
\begin{equation}
a=b=c=0;\qquad d\neq0.
\end{equation}
From these equations, we obtain the critical parameters, $S_{c}$, $\sigma_{c}$
and $T_{c}$.

We now sketch these calculations. From the equation of state, with $\eta=0$,
we write%
\begin{equation}
S=\frac{-1+\exp\left(  M\right)  }{2+\exp\left(  M\right)  },\label{Seqstat}%
\end{equation}
where%
\begin{equation}
M=\frac{9}{4}\beta\left[  1+\frac{1}{2}n_{s}T\delta^{2}\right]  S+\frac{3}%
{4}\beta\delta\sigma,
\end{equation}
and we remark that we are assuming a positive anisotropy, $\delta>0$, and the
number of polymeric strands per molecule, $n_{s}>0$, is also a characteristic
positive parameter. We then have the equation of state%
\begin{equation}
\frac{9}{4}\beta\left[  1+\frac{1}{2}n_{s}T\delta^{2}\right]  S+\frac{3}%
{4}\beta\delta\sigma=\ln\frac{1+2S}{1-S}.
\end{equation}
Inserting the form of $S$, given by equation (\ref{s}), and expanding in
powers of $\Delta S$, we have%
\begin{align}
&\frac{9}{4}\beta\left[  1+\frac{1}{2}n_{s}T\delta^{2}\right]  S_{c}+\frac
{9}{4}\beta\left[  1+\frac{1}{2}n_{s}T\delta^{2}\right]  \,s+\frac{3}{4}%
\beta\delta\sigma
\nonumber \\ &
=\ln\frac{1+2S_{c}}{1-S_{c}}+\left[  \frac{2}{1+2S_{c}}+\frac{1}{1-S_{c}%
}\right]  \,s+
\nonumber \\ &
+\frac{1}{2}\left[  -\frac{4}{\left(  1+2S_{c}\right)  ^{2}}+\frac{1}{\left(
1-S_{c}\right)  ^{2}}\right]  \,s^{2}+O\left(  s^{3}\right)  .
\end{align}

At the critical point, we write%
\begin{align}
a&=\frac{9}{4}\beta_{c}\left[  1+\frac{1}{2}n_{s}T_{c}\delta^{2}\right]
S_{c}+\frac{3}{4}\beta_{c}\delta\sigma_{c}
\nonumber \\ & \quad
-\ln\frac{1+2S_{c}}{1-S_{c}%
}=0,\label{c1}%
\end{align}%
\begin{equation}
b=\frac{9}{4}\beta_{c}\left[  1+\frac{1}{2}n_{s}T_{c}\delta^{2}\right]
-\left[  \frac{2}{1+2S_{c}}+\frac{1}{1-S_{c}}\right]  =0\label{c2}%
\end{equation}
and%
\begin{equation}
c=-\frac{1}{2}\frac{4}{\left(  1+2S_{c}\right)  ^{2}}+\frac{1}{2}\frac
{1}{\left(  1-S_{c}\right)  ^{2}}=0.\label{c3}%
\end{equation}

From eq. (\ref{c3}), we have%
\begin{equation}
S_{c}=\frac{1}{4}.
\end{equation}
We then use eq. (\ref{c2}), and write the critical temperature,%
\begin{equation}
T_{c}=\frac{27}{32}\left[  1-\frac{27}{64}n_{s}\delta^{2}\right]  ^{-1}.
\end{equation}

This critical temperature increases with the parameter $\omega$, given by%
\[
\omega=n_{s}\delta^{2},
\]
which gauges the strength of the coupling, and diverges for $\omega
\rightarrow64/27=2.3703...$, which is an indication that the contact with the
experimental situation is restricted to relatively small values of $\omega$.
We can also write an expression for the stress at the critical point,%
\[
\delta\sigma_{c}=\frac{4}{3}T_{c}\left[  \ln2-\frac{2}{3}\right]  =\left(
0.03531...\right)  T_{c},
\]
which is a linear function of $T_{c}$. We remark that this is a positive and
relatively small stress field, and that it does not make physical sense for
$\omega>64/27$.

As it is pointed out by Warner and Terentjev \cite{Warner2003}, the
engineering shear modulus of polymeric chains, at room temperatures, is of the
order of $10^{4}$ to $10^{6}$ GPa, which is at least five orders of magnitude
smaller than the corresponding shear modulus of usual solids. Therefore,
taking into account that $0<\delta<1$, at least for prolate elastomers, we
claim that physically realistic and accessible results will be restricted to
quite small values of the parameter $\omega=n_{s}\delta^{2}$. In some recent
numerical simulations for a uniaxial nematic elastomer on a lattice, Pasini
and coworkers \cite{Pasini2005}\ assumed that $n_{s}\approx0.3$. In the next
Section, we use a typical small value, $\omega=0.2$, to draw some graphs to
illustrate our main findings.

It is certainly interesting to make contact with the phenomenological
expansions of the free energy, which have been written by several authors, as
Selinger, Jeon, and Ratna \cite{Selinger2002,Selinger2004}. Let us then
consider the special uniaxial situation, and write an expansion of the
functional $g_{f}$, given by eq. (\ref{gfgen}), with $\eta=0$, as a power
series in the scalar order parameter,%
\begin{equation}
g_{f}=g_{f}\left(  0\right)  +A_{1}\,S+A_{2}\,S^{2}+A_{3}\,S^{3}%
+....,\label{gfLandau}%
\end{equation}
where the coefficients $A_{1}$, $A_{2}$, ..., depend on the thermodynamic
field variables, $T$ and $\sigma$, and on the model parameters. It is
straightforward to write the expressions of these coefficients. For example,
$A_{1}$ is given by%
\begin{equation}
A_{1}=\frac{3}{8}\mu_{s}\delta^{2}-\frac{3}{2}\frac{-1+\exp\left(  \frac{3}%
{4}\delta\sigma\right)  \left(  1+\frac{3}{4}\mu_{s}\delta^{2}\right)
}{2+\exp\left(  \frac{3}{4}\beta\delta\sigma\right)  }.
\end{equation}
From this equation, we see that $A_{1}=0$ for $\sigma=0$, regardless of the
value of the distortion, which confirms the existence of a disordered phase at
sufficiently high temperatures. It is not difficult to relate the coefficients
of the Landau expansion (\ref{gfLandau}) with the corresponding coefficients
of eq. (\ref{sexp}). In particular, $A_{3}\neq0$ indicates the characteristic
first-order nature of the nematic transition. Also, we can use the Landau
expansion to check and confirm the location of the critical point in the
$\sigma-T$ phase diagram.

\subsection{Biaxial transitions}

If the external stress is sufficiently large, and negative, with the usual
positive strain anisotropy, $\delta>0$, the nematic directors tend to be
parallel to the $x-y$ planes, which gives rise to biaxial ordering. In the
$\sigma-T$ phase diagram, we then predict the existence of a line of second
order transitions to a low-temperature biaxial phase.

This second-order line in the $\sigma-T$ plane can be located from an analysis
of the general equations of state for small values of the order parameter
$\eta$. Taking into account equations (\ref{S}), (\ref{eta}), and
(\ref{lambda}), we write the expansions%
\begin{equation}
S=S_{0}+S_{2}\eta^{2}+O\left(  \eta^{4}\right)  ,
\end{equation}
and%
\begin{equation}
1=E_{1}+E_{2}\eta^{2}+O\left(  \eta^{4}\right)  ,
\end{equation}
where the coefficients $S_{0}$, $S_{2}$, $E_{1}$, $E_{2}$, ...., depend on
stress, temperature, and the parameters of the model system.

The line of stability of the biaxial solution, $\eta^{2}\neq0$, comes from the
equation%
\begin{equation}
1=E_{1},\label{stability}%
\end{equation}
which leads to the expressions%
\begin{equation}
\frac{9}{4}\beta-2=\frac{2S_{0}+1}{1-S_{0}},
\end{equation}
and%
\begin{equation}
\frac{2S_{0}+1}{1-S_{0}}=\exp\left[  \frac{9}{4}\beta S_{0}+\frac{3}{4}%
\beta\delta\sigma+\frac{9}{8}n_{s}\delta^{2}S_{0}\right]  ,
\end{equation}
where we have set $A=1$ and assumed that the shear modulus depends linearly on
temperature. It is then straightforward to obtain the limit of stability of
the biaxial solution,%
\begin{align}
\left(  \sigma\delta\right)  _{crit} &=\frac{4T}{3}\ln\left(  \frac{9}%
{4T}-2\right)
-\left(  1+\frac{1}{2}n_{s}\delta^{2}T\right)
\nonumber \\ \quad & \quad \times
\left(
3-4T\right).
\end{align}
At this line, we have $\eta=0$ and $S=S_{0}$, with%
\begin{equation}
S_{0}=1-\frac{4T}{3}.\label{S0}%
\end{equation}
Note the asymptotic value $\sigma\delta\rightarrow-\infty$ for $T\rightarrow
9/8=1.125$.

We now look at the condition $E_{2}=0$, which is associated with a quartic
term of a Landau expansion, so that the second-order biaxial transition is
unstable for $E_{2}<0$. In the $\sigma-T$ phase diagram, the quartic line
$E_{2}=0$ is given by%
\begin{align}
\left(  \delta\sigma\right)  _{stab}&=\frac{4T}{3}\ln\left[  \left(  1-\frac
{3}{8T}\right)  \frac{1}{\left(  1+\frac{1}{2}n_{s}\delta^{2}T\right)
}\right]
\nonumber \\ & \quad
-\left(  1+\frac{1}{2}n_{s}\delta^{2}T\right)  \left(  3-4T\right).
\end{align}
Conditions $E_{1}=1$ and $E_{2}=0$, which correspond to the intersection of
curves $\left(  \sigma\delta\right)  _{crit}$ and $\left(  \delta
\sigma\right)  _{stab}$ versus temperature, determine the location of a
tricritical point, which comes from the equation%
\begin{equation}
9-8T=\frac{8T-3}{2\left(  1+\frac{1}{2}n_{s}\delta^{2}T\right)  },
\end{equation}
so that we have the physical solution%
\begin{equation}
T_{tri}=
\frac{9 \, \omega - 24 + \sqrt{576 +240 \, \omega +81 \, \omega^2}}{ 16 \, \omega }
\end{equation}
where $\omega=n_{s}\delta^{2}$.

In figure 2, we draw $\left(  \sigma\delta\right)_{crit}$ and $\left(
\sigma\delta\right)  _{stab}$ as a function of temperature for $\omega
=n_{s}\delta^{2}=0.2$, which is a typical small value of the parameter $\omega$.
The intersection of the critical line ($E_{1}=1$, green) with the limit of
stability of the biaxial solution ($E_{2}=0$; blue, dot-dashed) defines a
tricritical point. A number of calculations \cite{Palffy1983,Frisken1987,
Gramsbergen1986}, including our own work for the
Maier-Saupe-Zwanzig lattice model \cite{Petri2016}, indicate that the
qualitative features of the stress-temperature phase diagram, for sufficiently
small values of the coupling $\omega$, are also present in the
field-temperature phase diagram of a uniaxial nematic system.%
\begin{figure}
[ptb]
\begin{center}
\includegraphics[width = 0.85\linewidth]{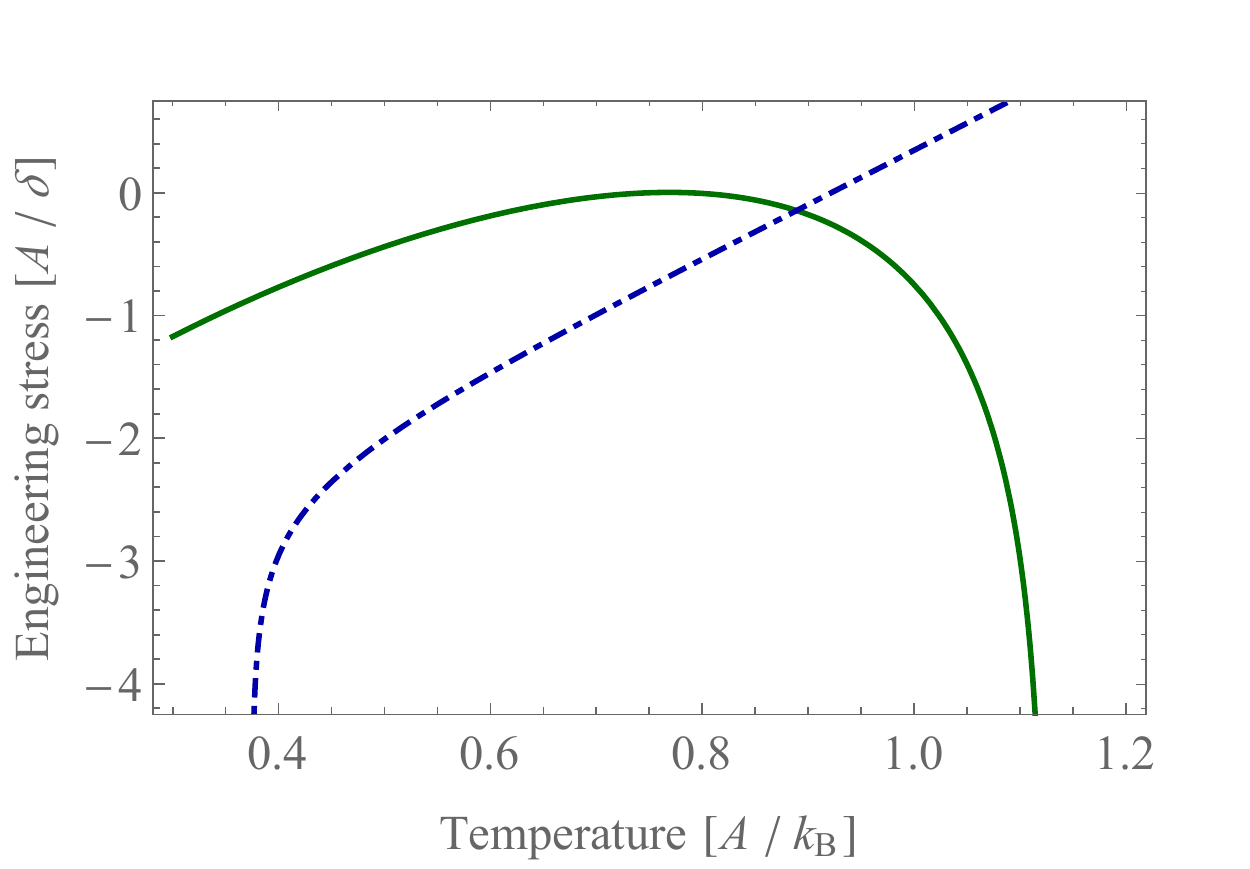}%
\caption{Critical line ($E_{1}=1$, green) of the biaxial solution, and limit
of stability of the biaxial solution ($E_{2}=0$; blue, dot-dashed) for a
typical paramater value, $\omega=n_{s}\delta^{2}=0.2$. The intersection
defines the tricritical point.}%
\end{center}
\end{figure}

\begin{figure}[ptb]
\begin{center}
\includegraphics[width=0.85\linewidth]{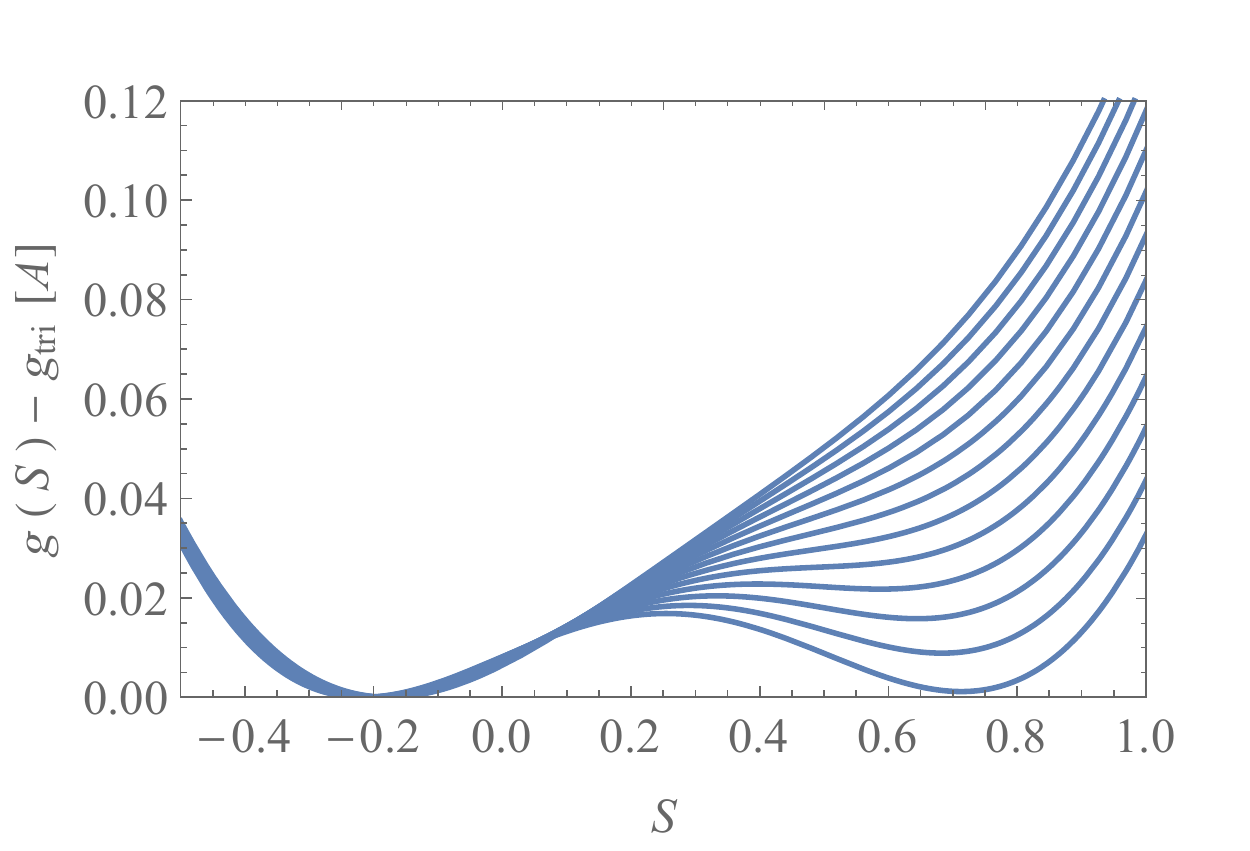}%
\caption{Free energy as a function of the nematic order parameter $S$ at the
tricritical point, for $\omega$ between $0.1$ (top curve) and $0.58$ (bottom
curve). The vertical axis is shifted so that $g(S_{0})=0$. }%
\label{fig:freeEnergy}
\end{center}
\end{figure}

In figure~\ref{fig:PhaseDiagram}a, we have drawn the stress-temperature
phase diagram for sufficiently small values of the coupling $\omega$. As this 
coupling parameter increases, the triple and tricritical points approach each
other, until they coalesce at $\omega=\omega_{top}$. We can describe this
topology change by considering the free energy, given by eq. (\ref{gfgen}), as
a function of $S$, \emph{at the tricritical point}. In figure~\ref{fig:freeEnergy}, we
show a plot of $S$ versus the free energy $g$, for $\omega$ between $0.1 $
(top curve) and $0.58$ (bottom curve), with the vertical axis shifted so that
$g(S_{0})=g_{\text{tri}}=0$. Note that the solution $S=S_{0}$ in eq. (\ref{S0})
becomes metastable at the threshold coupling $\omega=\omega_{top}$. The
value of the threshold coupling $\omega_{top}$ comes from the solutions of
equations $g(S)=g(S_{0})$ and the equation of state (\ref{Seqstat}), for $S$
($\neq S_{0}$) and $\omega$, at the tricritical point. Our numerical calculations
yield $\omega_{top}\approx0.59$ and $S_{top}\approx0.72$. For
$\omega>\omega_{top}$, the stress-temperature phase diagram consists of
just a first and a second-order transition lines. The first-order transition ends at
a simple critical point, and the second-order critical line ends at this first-order
border (see (figure~\ref{fig:PhaseDiagram}b)).%

The existence of a biaxial phase as well as the asymptotic limit of the
second-order transition to a paranematic region can be easily checked in the
infinite stress limit. From the equations of state, for $\sigma\delta
\rightarrow-\infty$, with $T\neq0$, we obtain the limiting values%
\begin{equation}
S\rightarrow\frac{-\exp\left(  -\frac{3}{4}\beta S\right)  \cosh\left(
\frac{3}{4}\beta\eta\right)  }{2\exp\left(  -\frac{3}{4}\beta S\right)
\cosh\left(  \frac{3}{4}\beta\eta\right)  }=-\frac{1}{2}%
\end{equation}
and%
\begin{equation}
\eta\rightarrow\frac{3\exp\left(  -\frac{3}{4}\beta S\right)  \sinh\left(
\frac{3}{4}\beta\eta\right)  }{2\exp\left(  -\frac{3}{4}\beta S\right)
\cosh\left(  \frac{3}{4}\beta\eta\right)  }=\frac{3}{2}\tanh\left(  \frac
{3}{4}\beta\eta\right)  .
\end{equation}
This second equation already indicates the characteristic up-down symmetry of
the biaxial phase in this discrete model system. There is a second order
transition at the temperature $T_{c}=9/8=1.125$, which is the previously found
asymptotic value of the critical temperature. Using the standard notation for
the nematic order parameter, as in equation (\ref{Q}), we have%
\begin{align}
Q&=\left(
\begin{array}
[c]{ccc}%
-\frac{1}{2}\left(  S+\eta\right)  & 0 & 0\\
0 & -\frac{1}{2}\left(  S-\eta\right)  & 0\\
0 & 0 & S
\end{array}
\right)
\nonumber \\ &
=\left(
\begin{array}
[c]{ccc}%
\frac{1}{4}-\frac{1}{2}\eta & 0 & 0\\
0 & \frac{1}{4}+\frac{1}{2}\eta & 0\\
0 & 0 & -\frac{1}{2}%
\end{array}
\right)  ,
\end{align}
which is a biaxial tensor except at the trivial limits $\eta=0$ (at the
second-order transition) and $\eta=\pm3/2$ (at the ground state).

\section{Conclusions}

We have analyzed the global phase diagram of a fully-connected version of a
Maier-Saupe-Zwanzig lattice model with the inclusion of couplings to an
elastic strain field. This is perhaps the simplest model system to be amenable
to standard statistical mechanics calculations to investigate elastic effects
on a nematic phase transition. We show the presence of uniaxial and biaxial
nematic structures, depending on temperature $T$ and on the applied stress
$\sigma$. If the applied stress favors uniaxial orientation, we obtain a
first-order boundary, along which there is a coexistence of two uniaxial
nematic phases, and which ends at a simple critical point. We locate this
critical point in terms of the model parameters. This picture is in
qualitative agreement with the experimental findings for the behavior of the
nematic order parameter of elastomers in a stress field. However, in the
presence of a compressive stress, which favors biaxial orientation, we show
the existence of a biaxially ordered region. Depending on the strength of the
couplings, there is a first-order boundary, but it ends at a tricritical
point, beyond which there is a continuous transition to a biaxially ordered
structure. We point out the analogy with the behavior of a nematic system in
the presence of an applied magnetic (electric) field, depending on the
strength of the field and on the sign of the anisotropy. Some of our analytic
results, for the critical and tricritical points, for example, may turn out to
be a helpful guide to experimental work. In the future, we plan to perform
Monte Carlo simulations to make contact with published numerical work
\cite{Pasini2005}, and to investigate the persistence of these qualitative
results in a scenario of short-range interactions.

\acknowledgments

We thank professor Mark Warner for useful conversations. S.R. Salinas
thanks the support of the Brazilian foundations CNPq and FAPESP. A. Petri
thanks the hospitality of the Institute of Physics of the University of S\~ao Paulo
and the South American Institute for Fundamental Research, S\~ao Paulo,
Brazil. D. B. Liarte was supported by NSF DMR-1719490.


\begin{thebibliography}{99}                                                                                               %
\bibitem {Warner2003}M. Warner and M. Terentjev, Liquid Crystal Elastomers,
Oxford University Press, Oxford, 2003.

\bibitem {deGennes1975}P. G. de Gennes, C. R. Seances Acad. Sci., Ser. A
\textbf{281}, 101, 1975.

\bibitem {deGennesProst1993}P. G. de Gennes and J. Prost, The Physics of
Liquid Crystals, Clarendon Press, Oxford, 1993.

\bibitem {deGennesOkumura2003}P. G. de Gennes and Ko Okumura, Europhys. Lett.
\textbf{50}, 513, 2000.

\bibitem {Fridrikh1997}S. V. Fridrikh and E. M. Terentjev, Phys. Rev. Lett.
\textbf{79}, 4661, 1997.

\bibitem {Yu1998}Y.-K. Yu, P. L. Taylor and E. M. Terentjev, Phys. Rev. Lett.
\textbf{81}, 128, 1998.

\bibitem {Selinger2002}J. V. Selinger, H. G. Jeon, and B. R. Ratna, Phys. Rev.
Lett. \textbf{89}, 225701, 2002.

\bibitem {Selinger2004}J. V. Selinger and B. R. Ratna, Phys. Rev. E
\textbf{70}, 041707, 2004.

\bibitem {Fridrich2006}S. V. Fridrich and E. M. Terentjev, Phys. Rev. E
\textbf{60}, 1847, 1999.

\bibitem {Petridis2006}L. Petridis and E. M. Terentjev, Phys. Rev. E
\textbf{74}, 051707, 2006.

\bibitem {Zhu2011}Wei Zhu, Michael Shelley, and Peter Palffy-Muhoray, Phys.
Rev. E \textbf{83}, 051703, 2011.

\bibitem {Liarte2011}Danilo B. Liarte, Silvio R. Salinas, and Carlos S. O.
Yokoi, Phys. Rev. E \textbf{84}, 011124, 2011.

\bibitem {Liarte2013}Danilo B. Liarte, Phys. Rev. E \textbf{88}, 062144, 2013.

\bibitem {doCarmo2011}E. do Carmo, A. P. Vieira, S. R. Salinas, Phys. Rev. E
\textbf{83}, 011701, 2011.

\bibitem {Liarte2014}D. B. Liarte and S. R. Salinas, Elementary statistical
models for nematic transitions in liquid-crystalline systems, in Perspectives
and Challenges in Statistical Physics for the Next Decade, G. M. Viswanathan,
E. P. Raposo, M. G. E. da Luz, editors, World Scientific, Singapore, 2014, p. 64.

\bibitem {Petri2016}A. Petri and S. R. Salinas, Field-induced uniaxial and
biaxial nematic phases in the Meier-Saupe-Zwanzig lattice model, unpublished.

\bibitem {Ye2007}F. Ye, R. Muckhopadhyay, O. Stenull, and T. C. Lubensky,
Phys. Rev. Lett. \textbf{98}, 147801, 2007.

\bibitem {Ye2009}F. Ye and T. C. Lubensky, J. Phys. Chem. B \textbf{113},
3853, 2009.

\bibitem {Kuepfer1991}J. K\"upfer and H. Finkelmann, Makromol. Chem. Rapid
Commun. \textbf{12}, 717, 1991.

\bibitem {Nascimento2016}E. S. Nascimento, A. P. Vieira, and S. R. Salinas,
Braz. J. Phys. \textbf{46}, 664 (2016).

\bibitem {Hornreich1985}R. M. Hornreich, Phys. Lett. A \textbf{109}, 232, 1985.

\bibitem {Pasini2005}P. Pasini, G. Skacej, and C. Zannoni, Chem. Phys. Lett
413, 463, 2005.

\bibitem {Palffy1983}P. Palffy-Muhoray and D. A. Dunmur, Mol. Cryst. Liq.
Cryst. \textbf{97}, 337, 1983.

\bibitem {Frisken1987}B. J. Frisken, B. Bergersen, and P. Palffy-Muhoray, Mol.
Cryst. Liq. Cryst. \textbf{148}, 45, 1987.

\bibitem {Gramsbergen1986}E. F. Gramsbergen, L. Longa, and Wim H. de Jeu,
Phys. Repts. \textbf{135}, 195, 1986.
\end{thebibliography}
\end{document}